\documentclass[twocolumn,prl,preprintnumbers,showpacs,amsmath,amssymb]{revtex4}

\usepackage{graphicx}
\usepackage{dcolumn}
\usepackage{bm}

\begin{document}

\title [] {Strain driven onset of non-trivial topological insulating states
in Zintl Sr$_2X$ compounds ($X$=Pb, Sn)}

\author{Yan Sun$^1$}
\author{Xing-Qiu Chen$^1$}
\email[The corresponding author should be addressed:
]{xingqiu.chen@imr.ac.cn}
\author{Dianzhong Li$^1$}
\author{Cesare Franchini$^{1,2}$}
\author{Seiji Yunoki$^3$}
\author{Yiyi Li$^1$}
\author{Zhong Fang$^4$}

\affiliation{$^1$ Shenyang National Laboratory for Materials
Science, Institute of Metal Research, Chinese Academy of Sciences,
Shenyang 110016, China}

\affiliation{$^2$ Center for Computational Materials Science,
University of Vienna, Sensengasse 8, A-1090 Vienna, Austria }

\affiliation{$^3$ Computational Condensed Matter Physics Laboratory,
RIKEN ASI, Saitama 351-0198, Japan, and CREST, Japan Science and
Technology Agency (JST), Saitama 332-0012, Japan, Computational
Materials Science Research Team, RIKEN AICS, Hyogo 650-0047}

\affiliation{$^4$ Beijing National Laboratory for Condensed Matter
Physics, Institute of Physics, Chinese Academy of Sciences, Beijing,
100081, China}

\date{\today}

\begin{abstract}
We explore the topological behavior of the binary Zintl phase of the
alkaline earth metals based compounds Sr$_2$Pb and Sr$_2$Sn using
both standard and hybrid density functional theory. It is found that
Sr$_2$Pb lies on the verge of a topological instability which can be
suitably tuned through the application of a small uniaxial expansion
strain ($>$ 3\%). The resulting non-trivial topologically insulating
state display well-defined metallic states in the Sr$_{2}$Pb(010)
surface, whose evolution is studied as a function of the film
thickness.
\end{abstract}


\maketitle

Topological insulators\cite{1,2,3,4,5,7} (TI) are a new quantum
state of matter characterized by the existence of gapless surface
states sheltered against destructive scattering effects by
time-reversal symmetry. Since the discovery of two dimensional TI
behaviors of HgTe-based quantum wells \cite{26,27}, several families
of topological materials have been theoretically predicted and
experimentally realized afterwards.\cite{1,30} The variety of
insulating materials displaying topological features includes $\rm
Bi_xSb_{1-x}$ alloys\cite{28,5}, Bi$_2$Te$_3$, Bi$_2$Se$_3$ and
Sb$_2$Te$_3$ binary compounds \cite{7,72,29,30,73}, ternary
heavy-metals based compounds such as TlBiTe$_2$, TlBiSe$_2$
\cite{31,32,33}, and PbBi$_2$Se$_4$ \cite{34,35}, and ternary
rare-earth chalcogenides (LaBiTe$_3$) \cite{36}, and another
honeycomb-lattice type of ternary compounds (LiAuSe) \cite{Zhanghj}.
All these materials are characterized by a layered structure stacked
along the $c$-axis of the centrosymmetric hexagonal lattice, similar
to the structure of Bi$_2$Se$_3$. In addition, other ternary TIs
have been recently predicted such as the non-centrosymmetic cubic
zinc-blende HgTe-like phase (half-Heusler compounds \cite{37,38,39}
and I-III-VI2 and II-IV-V2 chalcopyrite semiconductors such as
AuTlTe$_2$ \cite{40,41}), and Ca$_3$NBi with a centrosymmetric
antiperovskite structure \cite{Sun}. Although a large number of
ternary TI have been found\cite{31,32,33,34,35,36,37,38,39,40,41},
up to now, binary TIs are only limited to three classes:
Bi$_{1-x}$Sb$_x$ alloys and the family of Bi$_2$Te$_3$, Bi$_2$Se$_3$
and Sb$_2$Te$_3$ as well as Ag$_2$Te \cite{42}. In searching for new
TI materials we have focused our attention on binary
heavy-element-based small band gap semiconductors. One of the
simplest way to realize a binary heavy-element-based closed-shell
semiconductor is $M_2X$ with an alkaline earth element (\emph{M} =
Mg, Ca, Sr, and Ba) which donates its \emph{s} valence electron to a
IV group element (\emph{X} = Si, Ge, Sn and Pb). These are the
so-called Zintl compounds.

\begin{figure}[hbt]
\begin{center}
\includegraphics[width=0.45\textwidth]{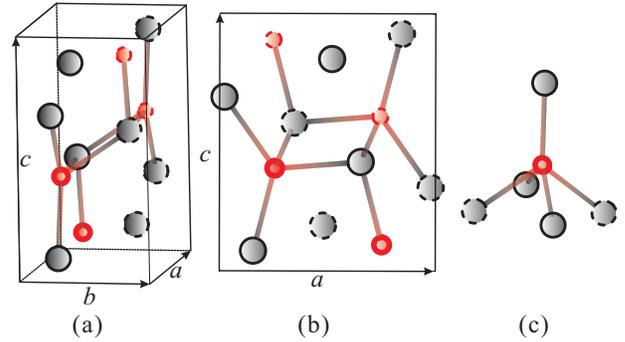}
\end{center}
\caption{(Color online) Structure representations of orthorhombic
(space group $Pnma$, No. 62) $M$$_2X$ compounds ($M$ = Mg, Ca, Sr
and Ba; $X$ = Si, Ge, Sn and Pb): (a) the unit cell (u.c.), (b) the
projection of the unit cell in the \emph{ac} plane, perpendicular to
the $b$-axis and (c) the local environment around the $X$ atom. The
large (gray) and small (red) balls denote the $M$ and $X$ atoms,
respectively. In addition, along the $b$-axis (y=1/4 and y=3/4) the
atoms can be arranged in two parallel planes shown in the dashed and
solid balls, respectively. $M$ occupied two inequivalent 4$c$ sites,
$M_1$ ($x_1$, 1/4, $z_1$) and $M_2$ ($x_2$, 1/4, $z_2$) whereas $X$
occupies the 4$c$ site ($x_3$, 1/4, $z_3$). For Sr$_2$Pb the
optimized atomic positions are Sr$_1$ (0.0214, 1/4, 0.683), Sr$_2$
(0.1580, 1/4, 0.073) and Pb (0.2496, 1/4, 0.3929). } \label{fig1}
\end{figure}

\begin{figure*}[hbt]
\begin{center}
\includegraphics[width=0.98\textwidth]{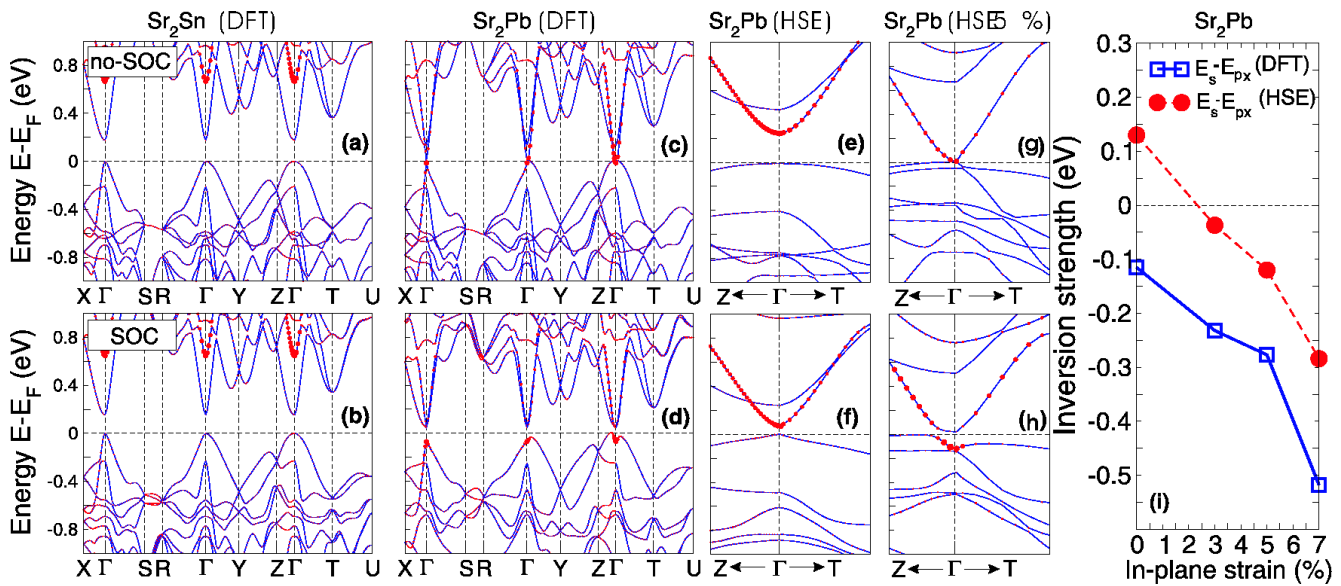}
\end{center}
\caption{(Color online) DFT and HSE electronic structure and band
inversion strength. Panels (a,b) and (c,d) show DFT calculated
electronic band structures for Sr$_2$Sn and Sr$_2$Pb, respectively.
Panels (e,f) and (g,h) reports HSE band structures around the
$\Gamma$ point along the Z-$\Gamma$-T directions for Sr$_2$Pb at
zero strain and under 5\% tensile strain in the {\em ac}-plane,
respectively. The upper and lower panels show the results calculated
without (a,c,e,g) and with (b,d,f,h) SOC, respectively. The solid
(red) circles denote the states with a predominant $s$-like
character. Panel (i) illustrates the comparison of the band
inversion strength between p$_x$- and $s$-orbitals at the $\Gamma$
point as a function of strain. Negative values denote the occurrence
of band inversion.} \label{fig2}
\end{figure*}

The Zintl compounds $M_2X$ crystallize in a simple orthorhombic
crystal structure \cite{Bruzzone,Eckerlin}, as illustrated in Fig.
\ref{fig1}. The $X$ atoms surround the $M$ atoms in a slightly
distorted trigonal-prismatic coordination. The coordination sphere
is augmented by the three $X$ atoms situated on the rectangular
faces of these prisms (tri-capped trigonal prism). Along the $a$
direction the prisms are condensed via common edges while they share
common triangular faces along the $b$ direction. The semiconducting
character of these compounds originates from their closed-shell
nature, with 6 valence electrons
(2$\times$\emph{s}$^2$+\emph{p}$^2$) per formula unit
[$(2M$$^{2+})X^{4-}$].

We have performed band structure calculations using the
Perdew-Burke-Ernzerhof\cite{pbe} (PBE) based standard and hybrid
(HSE\cite{hse}) density functional theory (DFT) as implemented in
the {\em Vienna Ab initio Simulation Package} (VASP)
\cite{vasp1,vasp2}, with the inclusion of relativistic spin-orbit
coupling (SOC) effects. Within the HSE method, the many body
exchange and correlation functional is constructed by mixing 25\%
exact Hartree-Fock exchange with 75\% PBE, and the long-range
Coulomb interaction is suitably screened according to the parameter
$\mu$ (here $\mu$ = 0.3 \AA\,$^{-1}$). By using the experimental
lattice constants we have relaxed all atomic positions using a very
dense k-points mesh (up to 4200 kpoints).

Our preliminary PBE-based research of possible topological features
in this class of materials revealed that two members of the Zintl
family (Sr$_2$Pb and Ba$_2$Pb) exhibit promising fingerprints of TI
behaviors whereas the remaining compounds are either trivial
insulator (Ca$_2$Ge, Sr$_2$Ge,Ca$_2$Sn and Sr$_2$Sn) or semimetal
(Ba$_2$Ge, Ca$_2$Pb and Ba$_2$Sn). Therefore in the following  we
focus our analysis on these two specific Zintl compounds: Sr$_2$Pb
and Sr$_2$Sn.

\begin{figure}[hbt]
\begin{center}
\includegraphics[width=0.48\textwidth]{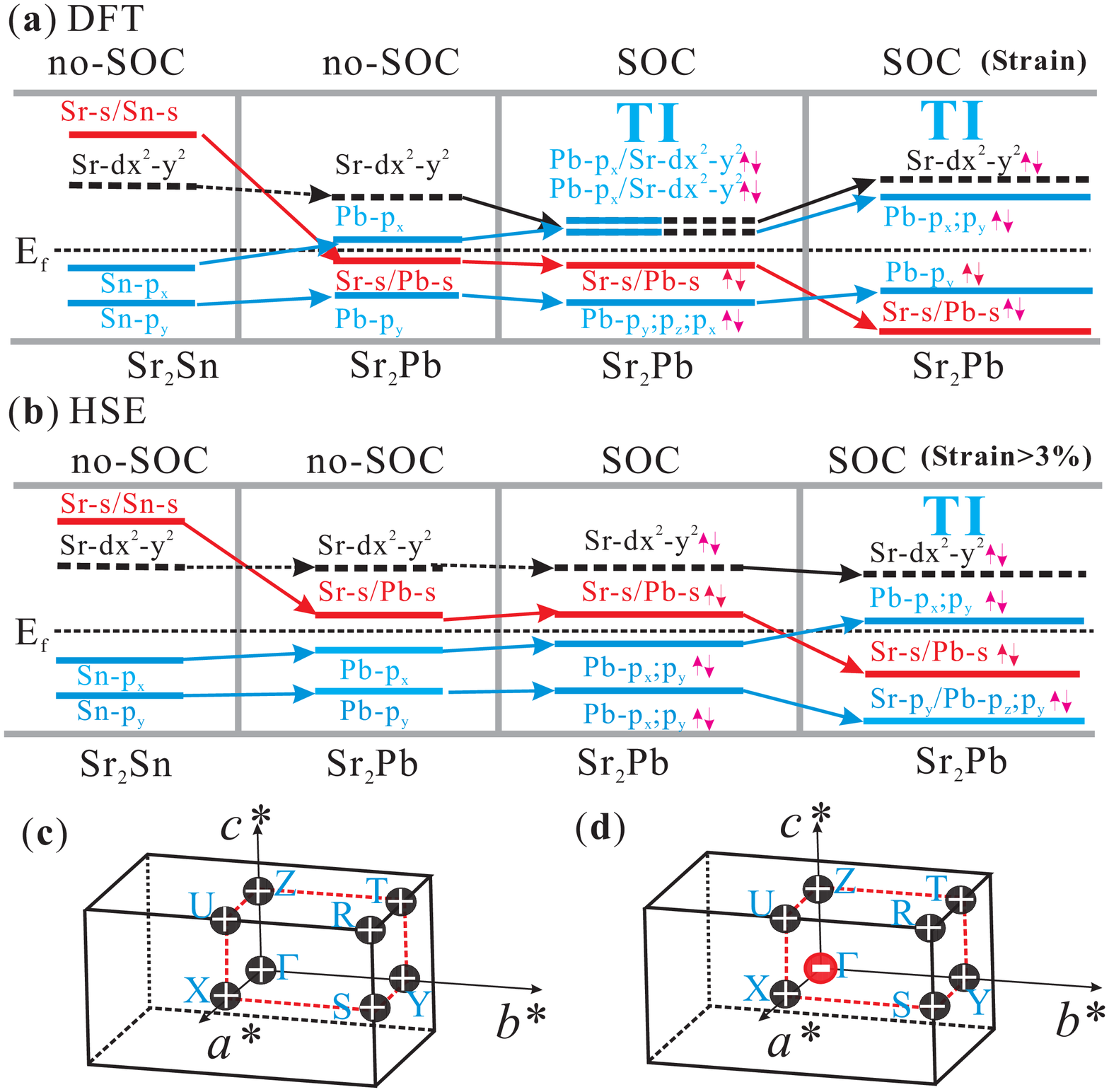}
\end{center}
\caption{(Color online) Evolution of atomic orbitals at the $\Gamma$
point from Sr$_2$Sn to  Sr$_2$Pb with and without SOC effects
included for both DFT (a) and HSE (b) calculations. The band
inversion can be observed by Pb (or Sr) $s$ and Pb $p_x$ orbitals.
In panels (c) and (d) we show the product of wave function parities
of the occupied bands  for eight time-reversal invariant momenta
(TRIM) in the Brillouin zone [$\Gamma$ (0,0,0), X ($\pi$,0,0), Y
(0,$\pi$,0), Z (0,0,$\pi$), S ($\pi$,$\pi$,0), T (0,$\pi$,$\pi$), U
($\pi$,0,$\pi$), and R ($\pi$,$\pi$,$\pi$)] of Sr$_2$Pb obtained
before and after band inversion, respectively.} \label{fig3}
\end{figure}

The DFT band structures with and without SOC effects are compared in
Fig. \ref{fig2}(a-d). The inclusion of SOC effects does not affect
the overall electronic character of Sr$_2$Sn  [Fig. \ref{fig2}
(a,b)] which remains semiconductor with a small bandgap at $\Gamma$
($\approx$ 0.15 eV) opened between occupied Sn $p_x$ states and
highly dispersive Sr $d$-like empty orbitals (mostly $d_{x^2-y^2}$).
By replacing Sn with the isoelectronic heavier element Pb the band
structure changes dramatically, as schematically depicted in
Fig.\ref{fig3} (a). Without SOC, Sr$_2$Pb displays a metallic
(gapless) state [Fig. \ref{fig2}(c)] whereas the inclusion of SOC
[Fig. \ref{fig2}(d)] opens a band gap of about 100 meV at $\Gamma$
(the indirect gap is about 50 meV), as a consequence of the
anti-crossing between the conduction band minimum (CBM) and valence
band maximum (VBM), a typical fingerprint of the SOC-induced
formation of topological insulating states. The major differences
between the electronic structure of Sr$_2$Sn and Sr$_2$Pb resides in
the orbital character of the valence and conduction bands near
$\Gamma$ as highlighted by the (red) solid circles in the band plot
of Fig.\ref{fig2}(a-d): in Sr$_2$Sn the states with a predominant
$s$-like character lie about 0.5 eV above the Fermi Level [see Fig.
\ref{fig2}(a,b)], whereas in Sr$_2$Pb these states are pushed down
in energy and eventually hybridize the VBM at $\Gamma$ thus inducing
the anti-crossing feature responsible for the opening of the gap and
the creation of a TI state. The downward shift of the $s$-states is
accompanied by a upward shift of the heavy metal $p$ bands which
ultimately intermix with the Sr $d$ states as schematized in Fig.
\ref{fig3} (a).

As recently reported by Zunger and coworkers\cite{Vidal}, the
identification of band-inverted TI\cite{26,27} on the basis of
conventional DFT may lead to false-positive assignment if the band
inversion strength is not large enough, due to the well documented
bandgap underestimation problem. For this reason we have revisited
the electronic dispersion of Sr$_2$Pb by HSE, which indeed yields a
substantially different physical picture, as illustrated in Fig.
\ref{fig2}(e,f). HSE finds a much larger bandgap (0.25 eV) which is
reduced to 0.13 eV with the inclusion of SOC effects, and, most
importantly, prevent the occurrence of band inversion between the
$p$$_x$-like and $s$-like states as schematically illustrated in
Fig. \ref{fig3}(b). The $s$ states remain well localized on the
bottom of the conduction band [Fig. \ref{fig2}(e,f)]: Sr$_2$Pb is
not a TI in its native phase. DFT wrongly stabilizes a spurious TI
solution because of the relatively small band inversion strength
[0.12 eV, see Fig.\ref{fig2}(a)]. Though HSE is expected to provide
a generally more accurate description of band dispersion in small
bandgap insulators, future optical experiment will be necessary in
order to validate our first principles findings.

To explore possible routes for designing a TI phase in the native
phase of Sr$_2$Pb, we carried out a series of calculations applying
strain, a gap-engineering technique which was successful when
applied to zero-gap semiconductors such as ternary Heusler
compounds\cite{38} and Ca$_3$NBi. \cite{Sun} We have chosen to study
the effect of uniaxial strain ($\epsilon$) in the $ac$-plane, by
leaving the $b$-axis unconstrained (free to relax) in order to
simulate at best the experimental condition for thin film growth.
The results, obtained by HSE and shown in Fig. \ref{fig2}(g and h),
indicate that for relatively small uniaxial strain larger than 3\%
Sr$_2$Pb can be tuned towards a TI phase. The role of SOC effects is
essential to open a small gap around $\Gamma$ and to induce an
inverted band order. The $s$-like states shift downward below the
Fermi level and becomes occupied and, simultaneously, the p$_x$-like
states become unoccupied and promoted at higher energy. This kind of
inverted band behavior can be ascribed to the fact that the
strain-induced expansion in the \emph{ac}-plane reduces the crystal
field effect, resulting in the less hybridization between \emph{s}-
and \emph{p}-like states and stabilizing the \emph{s}-like state at
lower energy, as evidenced in Fig. \ref{fig2}(e) and (g). In such a
way, the spin-orbit coupling strength is now enough to invert the
band order between \emph{s}-like and \emph{p}-like states at
$\Gamma$ (c.f., Fig. \ref{fig2}(f) and (h)). For larger strains up
to 7\% the TI state is preserved and further stabilized as inferred
by the evolution of the band inversion strength as a function of
strain reported in Fig. \ref{fig2}(i) at both, DFT and HSE levels.

Beside the band inversion, an alternative way to identify TI states
is the parity criteria proposed by Fu and Kane ~\cite{28}.
Considering that orthorhombic \emph{Pnma} posses the inversion
symmetry this criteria can be applied and will serve as a further
support for our analysis. The product of the parities of the Bloch
wave function for the occupied bands at all eight time reversal
invariant momenta (TRIM), illustrated in Fig. \ref{fig3}(c,d),
suggest that at six TRIMs (X, Y, Z, S, T and U) all bands share the
same doubly degenerate character, whereas the TRIM R is found to be
fourfold degenerate. Therefore all these seven TRIMs display a
positive ("+") global parity. At $\Gamma$ the situation is
different: the product of the parities is "+" and "-" depending on
whether or not the band inversion occurs, in consistency with the
HSE band structure interpretation [Fig. \ref{fig3}(c,d)]. We can
therefore trustfully conclude that distorted orthorhombic Sr$_2$Pb
is a topological non-trivial insulator with $\mathbb{Z}$$_2$ index
(1; 000). Strained-driven gap-engineering on Sr$_2$Sn dose not
results in any topological transition: Sr$_2$Sn remains a
conventional semiconductor.

\begin{figure}[hbt]
\begin{center}
\includegraphics[width=0.48\textwidth]{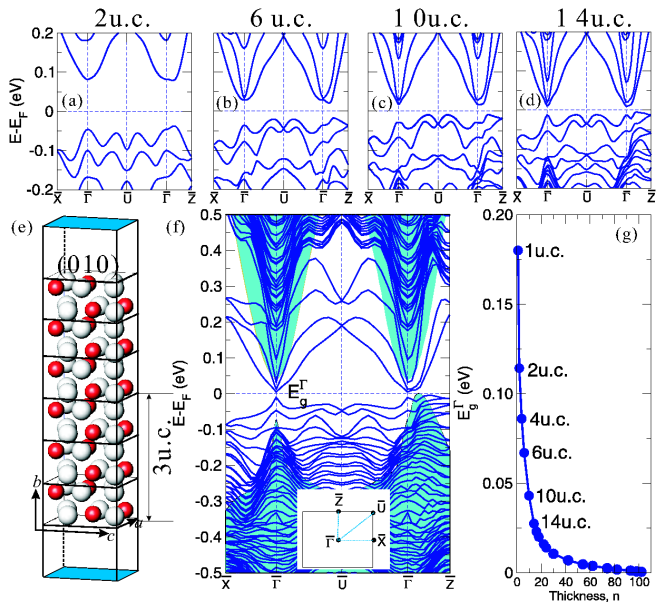}
\end{center}
\caption{(Color online) DFT Surface properties of strained
Sr$_2$Pb(010) ($\epsilon$ =5\%). (a--d) Evolution of the band
structures with slab thickness (e) structural model of the symmetric
slab adopted to simulate the Sr$_2$Pb(010) surface (the image
correspond to the 3 unit cells case, i.e. 6 layers per side. We
adopted a vacuum region of 30 \AA). (f) Band structure for a
thickness of 22 u.c. Shaded areas refer to bulk bands. (g)
Progressive closing of the gap at $\Gamma$ (E$_g^\Gamma$) as a
function of the slab thickness. Note that, due to the prohibitive
computational cost, for slab thickness larger than 34 u.c. ($>$ 400
atoms) the band gaps were calculated at $\Gamma$ only.} \label{fig4}
\end{figure}

After discussing the onset of topological features in the bulk phase
of distorted Sr$_2$Pb we turn our attention to the surface
properties focusing on the non-polar (010) termination. Considering
that both DFT and HSE lead to an essentially identical TI state in
strained Sr$_2$Pb and that SOC-HSE calculations for thick slabs are
computationally very demanding (if not prohibitive at all) we study
the surface band structure at DFT level only. The results on the
Sr$_2$Pb(010) surface are summarized in Fig. \ref{fig4}, which shows
that very thick slabs have topologically protected surface metallic
states which remain robust with increasing the film thickness, thus
corroborating the conclusion of the bulk parity analysis and the
inverted band order. The evolution of the band structure as a
function of the film thickness show a very peculiar behavior.
Already at low film thickness surface related states emerge in a
small energy window ($\pm$ 200 meV) around the Fermi level. This
leads to a quenching of the bandgap as compared to the bulk value
(130 meV). The bandgap is then progressively reduced by increasing
the slab thickness. Though for a thickness of 22 unit cell (u.c.)
[Fig. \ref{fig4}(f)] the gap is almost completely suppressed (10
meV), it is necessary to increase the thickness up to 100 u.c. in
order to fabricate a well-defined metallic film [Fig.
\ref{fig4}(g)]. At the critical thickness of 100 u.c. the lowest
conduction band and the highest valence band meet at the Fermi level
at $\Gamma$ and establishes a single-Dirac-cone-like metallic
surface state.

In conclusion, our computational study has disclosed the non-trivial
topological nature of the binary compound Sr$_2$Pb. The detailed
analysis of the bulk and surface structural and electronic
properties is of relevance for the design principles of TI and
provides helpful insight for tunability of TI states in trivial
insulator by gap-engineering techniques. We believe that our
findings will encourage immediate experimental investigations.

{\bf Acknowledgement}. The authors are grateful for supports from
the ``Hundred Talents Project'' of Chinese Academy of Sciences and
from the project of the NSFC of China (Grand Numbers: 51074151 and
51050110444) as well as Supercomputing Center of Chinese Academy of
Sciences (including Shenyang branch) and a local HPC cluster in the
Materials Process Simulation Division, IMR of CAS as well as the
Vienna Scientific Cluster (VSC) in Austria. F.Z. acknowledges the
supports from the 973 program of China (No. 2007CB925000).


\begin{thebibliography}{99}

\bibitem{1}  M. Z. Hasan, C. L. Kane,
   Rev. Mod. Phys. \textbf{82}, 3045 (2010).

\bibitem{2}
    J. E. Moore,
    Nature {\bf 464}, 194 (2010).

\bibitem{3} Zhang, S. -C.
    Physics {\bf 1}, 6 (2008)

\bibitem{4} D. Hsieh, D. Hsieh, Y. Xia, L. Wray, D. Qian, A. Pal,
J. H. Dil, F. Meier, J. Osterwalder, C. L. Kane, G. Bihlmayer, Y. S.
Hor, R. J. Cava and M. Z. Hasan. Science, \textbf{323}, 5916,
(2009).

\bibitem{5} D. Hsieh, D. Qian, L. Wray, Y. Xia, Y. S. Hor, R. J. Cava, and M. Z.
Hasan, Nature {\bf 452}, 970 (2008).

\bibitem{7} Y. Xia, D. Qian, D. Hsieh, L.Wray, A. Pal,
   H. Lin, A. Bansil, D. Grauer, Y. S. Hor, R. J. Cava
   and M. Z. Hasan,
    Nature Physics {\bf 5}, 398 (2009).

\bibitem{72}
D. Hsieh, J.W. McIver, D. H. Torchinsky, D. R. Gardner, Y. S. Lee,
and N. Gedik, Phys. Rev. Lett, {\bf 106}, 057401 (2011).

\bibitem{73}
Y. Xia, D. Qian, L. Wray, D. Hsieh, G. F. Chen, J. L. Luo, N. L.
Wang, and M. Z. Hasan, Phys. Rev. Lett, {\bf 103}, 037002 (2009).

\bibitem{26} B. A. Bernevig, T. L.Tughes and S.-C. Zhang,
    Science, \textbf{314}, 1757 (2006).

\bibitem{27} M. K\"onig, S. Wiedmann, C. Br\"une, A. Roth,
    H. Buhmann, L. Molenkamp, X. -L. Qi, and S. -C. Zhang,
    Science, \textbf{318}, 766 (2007).

\bibitem{30} Y. L. Chen, J. G. Analytis, J. -H. Chu, Z. K. Liu,
S.-K. Mo, X. -L. Qi, H. J. Zhang, D. H. Lu, X. Dai,
    Z. Fang, and S. -C. Zhang, Science, \textbf{325}, 178, (2009).

\bibitem{28}L. Fu and C.L. Kane, Phys. Rev. B, {\bf 76},
    045302 (2007).

\bibitem{29} H. J. Zhang,
C.-X. Liu, X.-L. Qi, X. Dai, Z. Fang,
   and S. -C. Zhang,
Nature Physics, \textbf{5}, 438 (2009).

\bibitem{31} B. Yan, C.-X. Liu, H.-J. Zhang, C.-Y. Yam,
    X.-L. Qi, T. Frauenheim, and S.-C. Zhang,
    Europhys. Lett. \textbf{90}, 37002 (2010).

\bibitem{32} Y. Chen, Z. Liu, J. Analytis, J. Chu, H.
    Zhang, S. Mo, R. Moore, D. Lu, I. Fisher, S. Zhang,
    arXiv:1006.3843v1 (2010).

\bibitem{33} T. Sato, K. Segawa, H. Guo, K. Sugawara,
S. Souma, T. Takahashi, Y. Ando, Phys. Rev. Lett. 105, 136802
(2010).

\bibitem{34} S.-Y. Xu, L.A. Wray, Y. Xia, R. Shankar,
    A. Petersen, A. Fedorov, H. Lin, A. Bansil, Y. S. Hor,
    R. J. Cava, and M. Z. Hasan, arXiv: 1007.5111 (2010).

\bibitem{35} H. Jin, J.-H. Song, A. J. Freeman and M. G. Kanatzidis,
    Phys. Rev. B, {\bf 83}, 041202(R) (2011).

\bibitem{36} B. H. Yan, H. -J. Zhang, C. -X. Liu, X. -L. Qi,
    T. Frauenheim, and S.-C. Zhang,
    Phys. Rev. B,{\bf 82}, 161108(R) (2010)

\bibitem{Zhanghj} H.-J. Zhang, S. Chadov, L. Muchler, B. Yan, X.-L. Qi,
J. K\"ubler, S.-C. Zhang, and C. Felser, Phys. Rev. Lett.,
\textbf{106}, 156402 (2011).

\bibitem{37} H. Lin, L. A. Wray, Y. Xia, S. Xu, S. Jia, R. J.
    Cava, A. Bansil, and M. Z. Hasan, Nature Mater. \textbf{9}, 546 (2010).

\bibitem{38} S. Chadov, X.
L. Qi, and et.al., Nature Mater. \textbf{9}, 541 (2010);

\bibitem{39} D. Xiao, Y. Y. Yao, W. Feng, J. Wen, W. Zhu, X.-Q.
    Chen, G. M. Stocks, and Z. Zhang,
    Phys. Rev. Lett. \textbf{105}, 096404 (2010).

\bibitem{40} M. Klintenberg,
    arXiv:1007.4838 (2010).

\bibitem{41} W. Feng, D. Xiao, J. Ding, and
    Y. Yao, Phys. Rev. Lett., {\bf 106}, 016402 (2011).

\bibitem{Sun} Y. Sun, X.-Q. Chen, S. Yunoki, D. Z. Li and Y. Y. Li,
    Phys. Rev. Lett., \textbf{105}, 216406 (2010).

\bibitem{42} W. Zhang, R. Yu, W. Feng, Y. Yao, H. Weng, X. Dai, and Z.
Fang, Phys. Rev. Lett., \textbf{106}, 156808 (2011).


\bibitem{Bruzzone} G. Bruzzone and E. Franceschi, J. Less-Comm Met.,
    {\bf 57}, 201-208 (1978).

\bibitem{Eckerlin}
   P.Eckerlin and E. Wolfel,
   Z. Anorg. Chem. \textbf{280}, 321 (1955).



\bibitem{pbe}
J. P. Perdew, K. Burke and M. Ernzerhof, Phys. Rev. Lett.{\bf 77},
3865 (1996).

\bibitem{hse}
J. Heyd, G. E. Scuseria, and M. Ernzerhof, J. Chem. Phys. {\bf 118},
8207 (2003).

\bibitem{vasp1} G. Kresse and J. Hafner, Phys.
    Rev. B {\bf 48}, 13115 (1993).

\bibitem{vasp2} G. Kresse and J. Furthmuller, Comput. Mater.
    Sci. {\bf 6}, 15 (1996).

\bibitem{Vidal} J. Vidal, X. Zhang, J.-W. Luo and A. Zunger,
arXiv:1101.3734 (2011).

\end{thebibliography}
\end{document}